# Constructing Infinitary Quotient-Inductive Types

Marcelo P. Fiore, Andrew M. Pitts, and S. C. Steenkamp[✉]

Department of Computer Science and Technology
University of Cambridge, Cambridge CB3 0FD, UK
`s.c.steenkamp@cl.cam.ac.uk`

**Abstract** This paper introduces an expressive class of quotient-inductive types, called QW-types. We show that in dependent type theory with uniqueness of identity proofs, even the infinitary case of QW-types can be encoded using the combination of inductive-inductive definitions involving strictly positive occurrences of Hofmann-style quotient types, and Abel's size types. The latter, which provide a convenient constructive abstraction of what classically would be accomplished with transfinite ordinals, are used to prove termination of the recursive definitions of the elimination and computation properties of our encoding of QW-types. The development is formalized using the Agda theorem prover.

## 1 Introduction

One of the key features of proof assistants based on dependent type theory such as Agda, Coq and Lean is their support for inductive definitions of families of types. Homotopy Type Theory [29] introduces a potentially very useful extension of the notion of inductive definition, the *higher inductive types* (HITs). To define an ordinary inductive type one declares how its elements are constructed. To define a HIT one not only declares element constructors, but also declares equality constructors in identity types (possibly iterated ones), specifying how the constructed elements and identities are to be equated. In this paper we work in a dependent type theory satisfying uniqueness of identity proofs (UIP), so that identity types are trivial in dimensions higher than one. Nevertheless, as Altenkirch and Kaposi [5] point out, HITs are still useful in such a one-dimensional setting. They introduce the term *quotient inductive type* (QIT) for this truncated form of HIT.

Figure 1 gives two examples of QITs, using Agda-style notation for dependent type theory; in particular, Set denotes a universe of types and ≡ denotes the identity type. The first example specifies the element and equality constructors for the type Bag $X$ of finite multisets of elements from a type $X$. The second example, adapted from [5], specifies the element and equality constructors for the type $\omega$Tree $X$ of trees whose nodes are labelled with elements of $X$ and that have unordered countably infinite branching. Both examples illustrate the nice feature

Finite multisets:
$$\begin{aligned}
&\text{data } \mathsf{Bag}(X : \mathsf{Set}) : \mathsf{Set} \text{ where}\\
&\quad [] : \mathsf{Bag}\, X\\
&\quad \_::\_ : X \to \mathsf{Bag}\, X \to \mathsf{Bag}\, X\\
&\quad \mathsf{swap} : (x\, y : X)(ys : \mathsf{Bag}\, X) \to x :: y :: ys \equiv y :: x :: ys
\end{aligned}$$

Unordered countably branching trees (elements of $\mathsf{isIso}\, f$ witness that $f$ is a bijection):
$$\begin{aligned}
&\text{data } \omega\mathsf{Tree}(X : \mathsf{Set}) : \mathsf{Set} \text{ where}\\
&\quad \mathsf{leaf} : \omega\mathsf{Tree}\, X\\
&\quad \mathsf{node} : X \to (\mathbb{N} \to \omega\mathsf{Tree}\, X) \to \omega\mathsf{Tree}\, X\\
&\quad \mathsf{perm} : (x : X)(f : \mathbb{N} \to \mathbb{N})(\_ : \mathsf{isIso}\, f)(g : \mathbb{N} \to \omega\mathsf{Tree}\, X) \to \\
&\qquad\qquad\qquad\qquad\qquad\qquad\qquad \mathsf{node}\, x\, g \equiv \mathsf{node}\, x\, (g \circ f)
\end{aligned}$$

**Figure 1.** Two examples of QITs

of QITs that users only have to specify the particular identifications between data needed for their applications. Thus the standard property of equality that it is an equivalence relation respecting the constructors is inherited by construction from the usual properties of identity types, without the need to say so in the declaration of the QIT.

The second example also illustrates a more technical aspect of QITs, that they enable constructive versions of structures that classically use non-constructive choice principles. The first example in Figure 1 only involves element constructors of finite arity ($[]$ is nullary and $x ::\_$ is unary) and consequently $\mathsf{Bag}\, X$ is isomorphic to the type obtained from the ordinary inductive type of finite lists over $X$ by quotienting by the congruence generated by swap. Of course this assumes, as we do in this paper, that the type theory comes with Hofmann-style *quotient types* [18, Section 3.2.6.1]. By contrast, the second example in the figure involves an element constructor with countably infinite arity. So if one first forms the ordinary inductive type of *ordered* countably branching trees (by dropping the equality constructor perm from the declaration) and then quotients by a suitable relation to get the equalities specified by perm, one needs the axiom of countable choice to be able to lift the node element constructor to the quotient; see [5, Section 2.2] for a detailed discussion. The construction of the Cauchy reals as a higher inductive-inductive type [29, Section 11.3] provides a similar, but more complicated example where use of countable choice is avoided. Such examples have led to the folklore that as far as constructive type theories go, infinitary QITs are more expressive than the combination of ordinary inductive (or inductive-recursive, or inductive-inductive) types with quotient types. In this paper we use Abel's *sized types* [2] to show that, for a wide class of QITs, this view is not justified. Thus we make two main contributions:

First we define a family of QITs called *QW-types* and give elimination and computation rules for them (Section 2). The usual W-types of Martin-Löf [22] are inductive types giving the algebraic terms over a possibly infinitary signature.

One specifies a QW-type by giving a family of equations between such terms. So such QITs give initial algebras for possibly infinitary algebraic theories. As we indicate in Section 3, they can encode a very wide range of examples of possibly infinitary quotient-inductive types, namely those that do not involve constructors taking previously constructed equalities as arguments (so do not cover the infinitary extension of the very general scheme considered by Dybjer and Moeneclaey [12]). In set theory with the Axiom of Choice (AC), QW-types can be constructed simply as Quotients of the underlying W-type—hence the name.

Secondly, we prove that contrary to expectation, without AC it is still possible to construct QW-types using quotients, but not simply by quotienting a W-type. Instead, the type to be quotiented and the relation by which to quotient are given simultaneously by definitions that refer to each other. Thus our construction (in Section 4) involves *inductive-inductive* definitions [15]. The elimination and computation functions which witness that the quotiented type correctly represents the required QW-type are defined recursively. In order to prove that our recursive definitions terminate we combine the use of inductive definitions involving strictly positive occurrences of quotient types with sized types (currently, we do not know whether it is possible to avoid sizing in favour of, say, a suitable well-founded termination ordering). Sized types provide a convenient constructive abstraction of what classically would be accomplished with sequences of transfinite ordinal length.

**The type theory in which we work**

To present our results we need a version of Martin-Löf Type Theory with (1) uniqueness of identity proofs, (2) quotient types and hence also function extensionality, (3) inductive-inductive datatypes (with strictly positive occurrences of quotient types) and (4) sized types. Lean 3 provides (1) and (2) out of the box, but also the Axiom of Choice, unfortunately. Neither it, nor Coq provide (3) and (4). Agda provides (1) via unrestricted dependent pattern-matching, (2) via a combination of postulates and the rewriting mechanism of Cockx and Abel [8], (3) via its very liberal mechanism for mutual definitions and (4) thanks to the work of Abel [2]. Therefore we make use of the type theory implemented by Agda (version 2.6.0.1) to give formal proofs of our results. The Agda code can be found at DOI: 10.17863/CAM.48187. In this paper we describe the results informally, using Agda-style notation for dependent type theory. In particular we use Set to denote the universe at the lowest level of a countable hierarchy of (Russell-style) universes. We also use Agda's convention that an implicit argument of an operation can be made explicit by enclosing it in {braces}.

*Acknowledgement* We would like to acknowledge the contribution Ian Orton made to the initial development of the work described here. He and the first author supervised the third author's Master's dissertation *Quotient Inductive Types: A Schema, Encoding and Interpretation*, in which the notion of QW-type (there called a $W^+$-type) was introduced.

## 2  QW-types

We begin by recalling some facts about types of well-founded trees, the W-types of Martin-Löf [22]. We take *signatures* to be elements of the dependent product

$$\mathsf{Sig} = \sum A : \mathsf{Set}, (A \to \mathsf{Set}) \tag{1}$$

So a signature is given by a pair $\Sigma = (A, B)$ consisting of a type $A : \mathsf{Set}$ and a family of types $B : A \to \mathsf{Set}$. Each such signature determines a polynomial endofunctor [1, 16] $\mathsf{S}\{\Sigma\} : \mathsf{Set} \to \mathsf{Set}$ whose value at $X : \mathsf{Set}$ is the following dependent product

$$\mathsf{S}\{\Sigma\}X = \sum a : A, (B\,a \to X) \tag{2}$$

An $\mathsf{S}$-*algebra* is by definition an element of the dependent product

$$\mathsf{Alg}\{\Sigma\} = \sum X : \mathsf{Set}, (\mathsf{S}\,X \to X) \tag{3}$$

S-algebra morphisms $(X, s) \to (X', s')$ are given by functions $h : X \to X'$ together with an element of the type

$$\mathsf{isHom}\,h = (a : A)(b : B\,a \to X) \to s'(a, h \circ b) \equiv h(s(a,b)) \tag{4}$$

Then the W-type $\mathsf{W}\{\Sigma\}$ determined by $\Sigma$ is the underlying type of an initial S-algebra. More generally, Dybjer [11] shows that the initial algebra of any non-nested, strictly positive endofunctor on $\mathsf{Set}$ is given by a W-type; and Abbott, Altenkirch, and Ghani [1] extend this to the case with nested uses of W-types as part of their work on containers. (These proofs take place in extensional type theory [22], but work just as well in the intensional type theory with uniqueness of identity proofs and function extensionality that we are using here.)

More concretely, given a signature $\Sigma = (A, B)$, if one thinks of elements $a : A$ as names of operation symbols whose (not necessarily finite) arity is given by the type $B\,a : \mathsf{Set}$, then the elements of $\mathsf{W}\{\Sigma\}$ represent the closed algebraic terms (i.e. well-founded trees) over the signature. From this point of view it is natural to consider not only closed terms solely built up from operations, but also open terms additionally built up with variables drawn from some type $X$. As well as allowing operators of possibly infinite arity, we also allow terms involving possibly infinitely many variables (the second example in Figure 1 involves such terms). Categorically, the type $\mathsf{T}\{\Sigma\}X$ of such open terms is the free $\mathsf{S}$-algebra on $X$ and is another W-type, for the signature obtained from $\Sigma$ by adding the elements of $X$ as nullary operations. Nevertheless, it is convenient to give a direct inductive definition:

$$\begin{aligned}
\mathsf{data} : \mathsf{T}\{\Sigma : \mathsf{Sig}\}(X : \mathsf{Set}) : \mathsf{Set}\ \mathsf{where} & \\
\eta : X \to \mathsf{T}\,X & \\
\sigma : \mathsf{S}(\mathsf{T}\,X) \to \mathsf{T}\,X &
\end{aligned} \tag{5}$$

Given an $\mathsf{S}$-algebra $(Y, s) : \mathsf{Alg}\{\Sigma\}$ and a function $f : X \to Y$, the unique morphism of $\mathsf{S}$-algebras from the free $\mathsf{S}$-algebra $(\mathsf{T}\,X, \sigma)$ on $X$ to $(Y, s)$ has underlying function $\mathsf{T}\,X \to Y$ mapping each $t : \mathsf{T}\,X$ to the element $t \ggg\!\!= f$ in $Y$ defined[1] by recursion on the structure of $t$:

$$\begin{aligned}\eta\,x \ggg\!\!= f &= f\,x \\ \sigma(a,b) \ggg\!\!= f &= s(a, \lambda x \to b\,x \ggg\!\!= f)\end{aligned} \qquad (6)$$

As the notation suggests, $\ggg\!\!=$ is the Kleisli lifting operation ("bind") for a monad structure on $\mathsf{T}$; indeed, it is the free monad on the endofunctor $\mathsf{S}$.

The notion of "QW-type" that we introduce in this section is obtained from that of W-type by considering not only the algebraic terms over a given signature, but also equations between terms. To code equations we use a type-theoretic rendering of a categorical notion of equational system introduced by Fiore and Hur, referred to as *term equational system* [14, Section 2] and as *monadic equational system* [13, Section 5], here instantiated to free monads on signatures.

**Definition 1.** *A* system of equations *over a signature* $\Sigma : \mathsf{Sig}$ *is specified by*

- *a type* $E : \mathsf{Set}$ *(whose elements* $e : E$ *name the equations)*
- *a family of types* $V : E \to \mathsf{Set}$ *(*$V\,e : \mathsf{Set}$ *contains the variables used in the equation named* $e : E$*)*
- *for each* $e : E$*, elements* $l\,e$ *and* $r\,e$ *of type* $\mathsf{T}(V\,e)$*, the free* $\mathsf{S}$*-algebra on* $V\,e$ *(the terms with variables from* $V\,e$ *that are equated by the equation named* $e$*).*

*Thus a system of equations over* $\Sigma$ *is an element of the dependent product*

$$\mathsf{Syseq}\{\Sigma\} = \sum E : \mathsf{Set}, \sum V : (E \to \mathsf{Set}), \qquad (7)$$
$$((e : E) \to \mathsf{T}(V\,e)) \times ((e : E) \to \mathsf{T}(V\,e))$$

*An* $\mathsf{S}\{\Sigma\}$*-algebra* $\mathsf{S}\,X \to X$ satisfies *the system of equations* $\varepsilon = (E, V, l, r) : \mathsf{Syseq}\{\Sigma\}$ *if there is an element of type*

$$\mathsf{Sat}\{\varepsilon\}X = (e : E)(\rho : V\,e \to X) \to ((l\,e) \ggg\!\!= \rho) \equiv ((r\,e) \ggg\!\!= \rho) \qquad (8)$$

The category-theoretic view of QW-types is that they are simply $\mathsf{S}$-algebras that are initial among those satisfying a given system of equations:

**Definition 2.** *A* QW-type *for a signature* $\Sigma = (A, B) : \mathsf{Sig}$ *and system of equations* $\varepsilon = (E, V, l, r) : \mathsf{Syseq}\{\Sigma\}$ *is given by a type* $\mathsf{QW}\{\Sigma\}\{\varepsilon\} : \mathsf{Set}$ *equipped with an* $\mathsf{S}$*-algebra structure and a proof that it satisfies the equations*

$$\mathsf{qwintro} : \mathsf{S}(\mathsf{QW}) \to \mathsf{QW} \qquad (9)$$
$$\mathsf{qwequ} : \mathsf{Sat}\{\varepsilon\}(\mathsf{QW}) \qquad (10)$$

*together with functions that witness that it is the initial such algebra:*

$$\mathsf{qwrec} : (X : \mathsf{Set})(s : \mathsf{S}\,X \to X) \to \mathsf{Sat}\,X \to \mathsf{QW} \to X \qquad (11)$$
$$\mathsf{qwrechom} : (X : \mathsf{Set})(s : \mathsf{S}\,X \to X)(p : \mathsf{Sat}\,X) \to \mathsf{isHom}(\mathsf{qwrec}\,X\,s\,p) \qquad (12)$$
$$\mathsf{qwuniq} : (X : \mathsf{Set})(s : \mathsf{S}\,X \to X)(p : \mathsf{Sat}\,X)(f : \mathsf{QW} \to X) \to \qquad (13)$$
$$\mathsf{isHom}\,f \to \mathsf{qwrec}\,X\,s\,p \equiv f$$

---

[1] Note that the definition of $\ggg\!\!=$ depends on the $\mathsf{S}$-algebra structure $s$; in Agda we use *instance arguments* to hide this dependence.

Given the definitions of $\mathsf{S}\{\Sigma\}$ in (2) and $\mathsf{Sat}\{\varepsilon\}$ in (8), properties (9) and (10) suggest that a QW-type is an instance of the notion of quotient-inductive type [5] with element constructor qwintro and equality constructor qwequ. For this to be so, $\mathsf{QW}\{\Sigma\}\{\varepsilon\}$ needs to have the requisite dependently-typed elimination and computation[2] properties for these element and equality constructors. As Proposition 1 below shows, these follow from (11)–(13), because we are working in a type theory with function extensionality (by virtue of assuming quotient types). To state the proposition we need a dependent version of (6). For each

$$P : \mathsf{QW} \to \mathsf{Set}$$
$$p : (a : A)(b : B\,a \to \mathsf{QW}) \to ((x : B\,a) \to P(b\,x)) \to P(\mathsf{qwintro}(a,b)) \qquad (14)$$

type $X : \mathsf{Set}$, function $f : X \to \sum x : \mathsf{QW}, P\,x$ and term $t : \mathsf{T}(X)$, we get an element $\mathsf{lift}\,P\,p\,f\,t : P(t \ggg \mathsf{fst} \circ f)$ defined by recursion on the structure of $t$:

$$\begin{aligned}\mathsf{lift}\,P\,p\,f\,(\eta\,x) &= \mathsf{snd}(f\,x)\\ \mathsf{lift}\,P\,p\,f\,(\sigma(a,b)) &= p\,a\,(\lambda x \to b\,x \ggg (\mathsf{fst} \circ f))(\mathsf{lift}\,P\,p\,f \circ b)\end{aligned} \qquad (15)$$

**Proposition 1.** *For a QW-type as in the above definition, given $P$ and $p$ as in (14) and a term of type*

$$(e : E)(f : V\,e \to \sum x : \mathsf{QW}, P\,x) \to \mathsf{lift}\,P\,p\,f\,(l\,e) \equiv\equiv \mathsf{lift}\,P\,p\,f\,(r\,e) \qquad (16)$$

*there are elimination and computation terms:*

$\mathsf{qwelim} : (x : \mathsf{QW}) \to P\,x$
$\mathsf{qwcomp} : (a : A)(b : B\,a \to \mathsf{QW}) \to \mathsf{qwelim}(\mathsf{qwintro}(a,b)) \equiv p\,a\,b\,(\mathsf{qwelim} \circ b)$

*(Note that (16) uses McBride's heterogeneous equality type [23], which we denote by $\equiv\equiv$, because $\mathsf{lift}\,P\,p\,f\,(l\,e)$ and $\mathsf{lift}\,P\,p\,f\,(r\,e)$ inhabit different types, namely $P(l\,e \ggg \mathsf{fst} \circ f)$ and $P(r\,e \ggg \mathsf{fst} \circ f)$ respectively.)* □

The proof of the proposition can be found in the accompanying Agda code (DOI: 10.17863/CAM.48187).

So QW-types are in particular quotient-inductive types (QITs). Conversely, in the next section we show that a wide range of QITs can be encoded as QW-types. Then in Section 4 we prove:

**Theorem 1.** *In constructive dependent type theory with uniqueness of identity proofs (or equivalently the Axiom K of Streicher [27]) and universes with inductive-inductive datatypes [15] permitting strictly positive occurrences of quotient types [18] and sized types [2], for every signature and system of equations (Definition 1) there is a QW-type as in Definition 2.*

---

[2] We only establish the computation property up to propositional rather than definitional equality; so, using the terminology of Shulman [25], these are *typal* quotient-inductive types.

*Remark 1 (Free algebras).* Definition 2 defines QW-types as *initial* algebras. A corollary of Theorem 1 is that *free-algebras* also exist. In other words, given a signature $\Sigma$ and a type $X : \mathsf{Set}$, there is an $\mathsf{S}$-algebra

$$(\mathsf{F}\{\Sigma\}\{\varepsilon\}X \,,\, \mathsf{S}\{\Sigma\}(\mathsf{F}\{\Sigma\}\{\varepsilon\}X) \to \mathsf{F}\{\Sigma\}\{\varepsilon\}X)$$

satisfying a system of equations $\varepsilon$ and equipped with a function $X \to \mathsf{F}\{\Sigma\}\{\varepsilon\}X$, and which is universal among such $\mathsf{S}$-algebras. Thus $\mathsf{QW}\{\Sigma\}\{\varepsilon\}$ is isomorphic to $\mathsf{F}\{\Sigma\}\{\varepsilon\}\varnothing$, where $\varnothing$ is the empty datatype.

To see that such free algebras can be constructed as QW-types, given a signature $\Sigma = (A, B)$, let $\Sigma_X$ be the signature $(X \uplus A, B')$, where $X \uplus A$ is the coproduct datatype (with constructors $\mathsf{inl} : X \to X \uplus A$ and $\mathsf{inr} : A \to X \uplus A$) and where $B' : X \uplus A \to \mathsf{Set}$ maps each $\mathsf{inl}\,x$ to $\varnothing$ and each $\mathsf{inr}\,a$ to $B\,a$. Given a system of equations $\varepsilon = (E, V, l, r)$, let $\varepsilon_X$ be the system $(E, V, l_X, r_X)$ where for each $e : E$, $l_X\,e = l\,e \ggg \eta$ and $r_X\,e = r\,e \ggg \eta$ (using $\eta : V\,e \to \mathsf{T}\{\Sigma_X\}(V\,e)$ as in (5) and the $\mathsf{S}\{\Sigma\}$-algebra structure $s$ on $\mathsf{T}\{\Sigma_X\}(V\,e)$ given by $s(a,b) = \sigma(\mathsf{inr}\,a, b)$). Then one can show that the QW-type $\mathsf{QW}\{\Sigma_X\}\{\varepsilon_X\}$ is the free algebra $\mathsf{F}\{\Sigma\}\{\varepsilon\}X$, with the function $X \to \mathsf{F}\{\Sigma\}\{\varepsilon\}X$ sending each $x : X$ to $\mathsf{qwintro}(\mathsf{inl}\,x, \_) : \mathsf{QW}\{\Sigma_X\}\{\varepsilon_X\}$, and the $\mathsf{S}\{\Sigma\}$-algebra structure on $\mathsf{F}\{\Sigma\}\{\varepsilon\}X$ being given by the function sending $(a, b) : \mathsf{S}(\mathsf{QW}\{\Sigma_X\}\{\varepsilon_X\})$ to $\mathsf{qwintro}(\mathsf{inr}\,a, b)$.

*Remark 2 (Strictly positive equational systems).* A very general, categorical notion of equational system was introduced by Fiore and Hur [14, Section 3]. They regard any endofunctor $S : \mathsf{Set} \to \mathsf{Set}$ as a *functorial signature*. A *functorial term* over such a signature, $S \rhd G \vdash L$, is specified by another functorial signature $G : \mathsf{Set} \to \mathsf{Set}$ (the term's context) together with a functor $L$ from $S$-algebras to $G$-algebras that commutes with the forgetful functors to $\mathsf{Set}$. Then an *equational system* is given by a pair of such terms in the same context, $S \rhd G \vdash L$ and $S \rhd G \vdash R$ say. An $S$-algebra $s : S\,X \to X$ satisfies the equational system if $L(X, s)$ and $R(X, s)$ are equal $G$-algebras.

Taking the *strictly positive* endofunctors $\mathsf{Set} \to \mathsf{Set}$ to be the smallest collection containing the identity and constant endofunctors and closed under forming dependent products and dependent functions over fixed types then, as in [11] (and also in the type theory in which we work), up to isomorphism every such endofunctor is of the form $\mathsf{S}\{\Sigma\}$ for some signature $\Sigma : \mathsf{Sig}$. If we restrict attention to equational systems $S \rhd G \vdash L, R$ with $S$ and $G$ strictly positive, then it turns out that such equational systems are in bijection with the systems of equations from Definition 1, and the two notions of satisfaction for an algebra coincide in that case. (See our Agda development for a proof of this.) So Dybjer's characterisation of W-types as initial algebras for strictly positive endofunctors generalises to the fact that *QW-types are initial among the algebras satisfying strictly positive equational systems in the sense of Fiore and Hur.*

## 3 Quotient-inductive types

Higher inductive types (HITs) are originally motivated by their use in homotopy type theory to construct homotopical cell complexes, such as spheres, tori, and so on [29]. Intuitively, a higher inductive type is an inductive type with point constructors also allowing for path constructors, surface constructors, etc., which are represented as elements of (iterated) identity types. For example, the sphere is given by the HIT[3]:

$$
\begin{aligned}
&\mathsf{data}\ \mathsf{S}^2 : \mathsf{Set}\ \mathsf{where} \\
&\quad \mathsf{base} : \mathsf{S}^2 \\
&\quad \mathsf{surf} : \mathsf{refl} \equiv_{\mathsf{base} \equiv_{\mathsf{S}^2} \mathsf{base}} \mathsf{refl}
\end{aligned}
\tag{17}
$$

In the presence of the UIP axiom we will refer to HITs as *quotient inductive types* (QITs) [5], since all paths beyond the first level are trivial and any HIT is truncated to an h-set. We use the terms *element constructor* and *equality constructor* to refer to the point constructors and the only non-trivial level of path constructors.

We believe that QW-types can be used to encode a wide range of QITs: see Conjecture 1 below. As evidence, we give several examples of QITs encoded as QW-types, beginning with the two examples of QITs in Figure 1, giving the corresponding signature $(A, B)$ and system of equations $(E, V, l, r)$ as in Definition 2.

*Example 1 (Finite multisets).* The element constructors for finite multisets are encoded exactly as with a W-type: the constructors are $[]$ and $x :: \_$ for each $x : X$. So we take $A$ to be $\mathbb{1} \uplus X$, the coproduct of the unit type $\mathbb{1}$ (whose single constructor is denoted $\mathsf{tt}$) with $X$. The arity of $[]$ is zero, and the arity of each $x :: \_$ is one, represented by the empty type $\varnothing$ and unit type $\mathbb{1}$ respectively; so we take $B : A \to \mathsf{Set}$ to be the function $[\lambda\_ \to \mathbb{0} \mid \lambda\_ \to \mathbb{1}] : \mathbb{1} \uplus X \to \mathsf{Set}$ mapping $\mathsf{inl}\,\mathsf{tt}$ to $\varnothing$ and each $\mathsf{inr}\,x$ to $\mathbb{1}$.

The $\mathsf{swap}$ equality constructor is parameterised by elements of $E = X \times X$. For each $(x, y) : E$, $\mathsf{swap}\,x\,y$ yields an equation involving a single free variable (called $ys : \mathsf{Bag}\,X$ in Figure 1); so we take $V : E \to \mathsf{Set}$ to be $\lambda\_ \to \mathbb{1}$. Each side of the equation named by $\mathsf{swap}\,x\,y$ is coded by an element of $\mathsf{T}\{\Sigma\}(V(x,y)) = \mathsf{T}\{\Sigma\}(\mathbb{1})$. Recalling the definition of $\mathsf{T}$ from (5), the single free variable corresponds to $\eta\,\mathsf{tt} : \mathsf{T}\{\Sigma\}(\mathbb{1})$ and then the left-hand side of the equation is $\sigma(\mathsf{inr}\,x, (\lambda\_ \to \sigma(\mathsf{inr}\,y, (\lambda\_ \to \eta\,\mathsf{tt}))))$ and the right-hand side is $\sigma(\mathsf{inr}\,y, (\lambda\_ \to \sigma(\mathsf{inr}\,x, (\lambda\_ \to \eta\,\mathsf{tt}))))$.

So, altogether, the signature and system of equations for the QW-type corresponding to the first example in Figure 1 is:

$$
\begin{aligned}
A &= \mathbb{1} \uplus X & E &= X \times X \\
B &= [\lambda\_ \to \varnothing \mid \lambda\_ \to \mathbb{1}] & V &= \lambda\_ \to \mathbb{1} \\
l &= \lambda\,(x,y) \to \sigma(\mathsf{inr}\,x, (\lambda\_ \to \sigma(\mathsf{inr}\,y, (\lambda\_ \to \eta\,\mathsf{tt})))) \\
r &= \lambda\,(x,y) \to \sigma(\mathsf{inr}\,y, (\lambda\_ \to \sigma(\mathsf{inr}\,x, (\lambda\_ \to \eta\,\mathsf{tt}))))
\end{aligned}
$$

---

[3] The subscript on $\equiv$ will be treated as an implicit argument and omitted when clear.

*Example 2 (Unordered countably-branching trees).* Here the element constructors are leaf of arity zero and, for each $x : X$, node $x$ of arity $\mathbb{N}$. So we use the signature with $A = \mathbb{1} \uplus X$ and $B = [\lambda\_ \to \varnothing \mid \lambda\_ \to \mathbb{N}]$.

The perm equality constructor is parameterised by elements of

$$E = X \times \sum f : (\mathbb{N} \to \mathbb{N}),\, \mathsf{isIso}\, f$$

For each element $(x, f, i)$ of that type, perm $x\, f\, i$ yields an equation involving an $\mathbb{N}$-indexed family of variables (called $g : \mathbb{N} \to \omega\mathsf{Tree}\, X$ in Figure 1); so we take $V : E \to \mathsf{Set}$ to be $\lambda\_ \to \mathbb{N}$. Each side of the equation named by perm $x\, f\, i$ is coded by an element of $\mathsf{T}\{\Sigma\}(V(x, f, i)) = \mathsf{T}\{\Sigma\}(\mathbb{N})$. The $\mathbb{N}$-indexed family of variables is represented by the function $\eta : \mathbb{N} \to \mathsf{T}\{\Sigma\}(\mathbb{N})$ and its permuted version by $\eta \circ f$. Thus the left- and right-hand sides of the equation named by perm $x\, f\, i$ are coded respectively by the elements $\sigma(\mathsf{inr}\, x, \eta)$ and $\sigma(\mathsf{inr}\, x, \eta \circ f)$ of $\mathsf{T}\{\Sigma\}(\mathbb{N})$.

So, altogether, the signature and system of equations for the QW-type corresponding to the second example in Figure 1 is:

$$\begin{aligned}
A &= \mathbb{1} \uplus X & E &= X \times \sum f : (\mathbb{N} \to \mathbb{N}),\, \mathsf{isIso}\, f \\
B &= [\lambda\_ \to \varnothing \mid \lambda\_ \to \mathbb{N}] & V &= \lambda\_ \to \mathbb{N} \\
l &= \lambda\, (x, \_, \_) \to \sigma(\mathsf{inr}\, x, \eta) \\
r &= \lambda\, (x, f, \_) \to \sigma(\mathsf{inr}\, x, \eta \circ f)
\end{aligned}$$

That unordered countably-branching trees are a QW-type is significant since no previous work on various subclasses of QITs (or indeed QIITs [19, 10]) supports infinitary QITs [6, 26, 28, 12, 19, 10]. See Example 5 for another, more substantial infinitary QW-type. So this extension represents one of our main contributions. QW-types generalise prior developments; the internal encodings for particular subclasses of 1-HITs given by Sojakova [26] and Swan [28] are direct instances of QW-types, as the next two examples show.

*Example 3.* *W-suspensions* [26] are an instance of QW-types. The data for a W-suspension is: $A', C' : \mathsf{Set}$, a type family $B' : A' \to \mathsf{Set}$ and functions $l', r' : C' \to A'$. The equivalent QW-type is:

$$\begin{aligned}
A &= A' & E &= C' & l &= \lambda\, c \to \sigma((l'\, c), \eta) \\
B &= B' & V &= \lambda\, c \to (B'\, (l'\, c)) \times (B'\, (r'\, c)) & r &= \lambda\, c \to \sigma((r'\, c), \eta)
\end{aligned}$$

*Example 4.* The non-indexed case of *W-types with reductions* [28] are QW-types. The data of such a type is: $Y : \mathsf{Set}$, $X : Y \to \mathsf{Set}$ and a reindexing map $R : (y : Y) \to Xy$. The reindexing map identifies a term $\sigma\, (y, \alpha)$ with some $\alpha\, (R\, y)$ used to construct it. The equivalent QW-type is given by:

$$\begin{aligned}
A &= Y & E &= Y & l &= \lambda y \to \sigma\, (y, \eta) \\
B &= X & V &= X & r &= \lambda y \to \eta\, (R\, i)
\end{aligned}$$

*Example 5.* Lumsdaine and Shulman [21, Section 9] give an example of a HIT not constructible in type theory from only pushouts and $\mathbb{N}$. Their HIT $F$ can be thought of as a set of notations for countable ordinals. It consists of three point constructors: $0 : F$, $S : F \to F$, and $\mathsf{sup} : (\mathbb{N} \to F) \to F$, and five path constructors which are omitted here for brevity. It is inspired by the infinitary algebraic theory of Blass [7, Section 9] and hence it is not surprising that it can be encoded by a QW-type; the details can be found in our Agda code.

### 3.1  General QIT schemas

Basold, Geuvers, and van der Weide [6] present a schema (though not a model) for infinitary QITs that do not support conditional path equations. Constructors are defined by arbitrary polynomial endofunctors built up using (non-dependent) products and sums, which means in particular that parameters and arguments can occur in any order. They require constructors to be in uncurried form.

Dybjer and Moeneclaey [12, Sections 3.1 and 3.2] present a schema for finitary QITs that supports *conditional* path equations, where constructors are allowed to take inductive arguments not just of the datatype being declared, but also of its identity type. This schema can be generalised to infinitary QITs with conditional path equations. We believe this extension of their schema to be the most general schema for QITs. The schema requires all parameters to appear before all arguments, whereas the schema for regular inductive types in Agda is more flexible, allowing parameters and arguments in any order.

We wish to combine the schema for infinitary QITs of Basold, Geuvers, and van der Weide [6] with the schema for QITs with conditional path equations of Dybjer and Moeneclaey [12] to provide a general schema. Moreover, we would like to combine the arbitrarily ordered parameters and arguments of the former with the curried constructors of the latter in order to support flexible pattern matching.

For consistency with the definition of inductive types in Agda [9, equation (25) and figure 1] we will define strictly positive (i.e. polynomial) endofunctors in terms of strictly positive telescopes.

A telescope is given by the grammar:

$$\begin{aligned}\Delta ::= &\ \epsilon & \text{empty telescope} \\ | &\ (x : A)\Delta \quad (x \notin \mathrm{dom}(\Delta)) & \text{non-empty telescope}\end{aligned} \quad (18)$$

A telescope extension $(x : A)\Delta$ binds (free) occurrences of $x$ inside the tail $\Delta$. The type $A$ may contain free variables that are later bound by further telescope extensions on the left. A telescope can also exist in a context which binds any free variables not already bound in the telescope. Such a context is implicit in the following definitions. A function type $\Delta \to C$ from a telescope $\Delta$ to a type $C$ is defined as an iterated dependent function type by:

$$\begin{aligned}\epsilon \to C &\stackrel{\mathrm{def}}{=} C \\ (x : A)\Delta \to C &\stackrel{\mathrm{def}}{=} (x : A) \to (\Delta \to C)\end{aligned} \quad (19)$$

A *strictly positive* endofunctor on a variable $Y$ is presented by a strictly positive telescope

$$\Delta = (x_1 : \Phi_1(Y))(x_2 : \Phi_2(Y)) \cdots (x_n : \Phi_n(Y))\epsilon \tag{20}$$

where each type scheme $\Phi_i$ is described by a expression on $Y$ made up of $\Pi$-types, $\Sigma$-types, and any (previously defined "constant") types $A$ not containing $Y$, according to the grammar:

$$\Phi(Y), \Psi(Y) ::= \quad (y : A) \to \Phi(Y) \quad | \quad \Sigma\, p : \Phi(Y), \Psi(Y) \quad | \quad A \quad | \quad Y \tag{21}$$

For example, $\Delta \stackrel{\text{def}}{=} (x : X)(f : \mathbb{N} \to Y)\epsilon$ is the strictly positive telescope for the node constructor in Figure 1. In this instance, reordering $x$ and $f$ is permitted by exchange. Note that the variable $Y$ can never appear in the argument position of a $\Pi$-type.

Now it is possible to define the form of the endpoints of an equality (within the context of a strictly positive telescope), corresponding to the notion of an abstract syntax tree with free variables. With this intuition in mind, we can take the definition in Dybjer and Moeneclaey's presentation [12] of endpoints given by *point constructor patterns*:

$$l, r, p ::= \quad c_i\, k \quad | \quad y \tag{22}$$

Where $y : Y$ is in the context of the telescope for the equality constructor, and $k$ is a term built without any rule for $Y$, but which may use other point constructor patterns $p : Y$. (That is, any sub-term of type $Y$ must either be a variable $y : Y$ found in the telescope, or a constructor for $Y$ applied to further point constructor patterns and earlier defined constants. It could not, for instance, use the function application rule for $Y$ with some function $g : M \to Y$, not least since such functions cannot be defined before defining $Y$.) Note that this exactly matches the type $\mathsf{T}$ in (5).

Basold, Geuvers, and van der Weide's presentation has a sightly more general notion of *constructor term* [6, Definition 6] (Dybjer and Moeneclaey's presentation [12] has more restricted telescopes). It is defined by rules which operate in the context of a strictly positive (polynomial) telescope and permit use of its bound variables, and the use of constructors $c_i$, but not any other rules for $Y$. We take the dependent form of their rules for products and functions. Note that these rules do not allow the use of terms of type $\equiv_Y$ in the endpoints.

As with inductive types, the element constructors of QITs are specified by strictly positive telescopes. The equality constructors also permit *conditions* to appear in strictly positive positions, where $l$ and $r$ are constructor terms according to grammar (22):

$$\Phi(Y), \Psi(Y) ::= (\text{same grammar as in (21)}) \mid l \equiv_Y r \tag{23}$$

**Definition 3.** *A* QIT *is defined by a list of named element constructors and equality constructors:*

$$
\begin{aligned}
&\text{data } \mathsf{Y} : \mathsf{Set} \text{ where} \\
&\qquad \mathsf{c}_1 : \Delta_1 \to \mathsf{Y} \\
&\qquad \vdots \\
&\qquad \mathsf{c}_n : \Delta_n \to \mathsf{Y} \\
&\qquad \mathsf{p}_1 : \Theta_1 \to l_1 \equiv_{\mathsf{Y}} r_1 \\
&\qquad \vdots \\
&\qquad \mathsf{p}_m : \Theta_m \to l_m \equiv_{\mathsf{Y}} r_m
\end{aligned}
$$

*where $\Delta_i$ are strictly positive telescopes on $\mathsf{Y}$ according to* (21), *and $\Theta_j$ are strictly positive telescopes on $\mathsf{Y}$ and $\equiv_{\mathsf{Y}}$ in which conditions may also occur in strictly positive positions according to* (23).

QITs without equality constructors are inductive types. If none of the equality constructors contain $Y$ in an argument position then it is called *non-recursive*, otherwise it is called *recursive* [6]. If none of the equality constructors contain an equality in $Y$ then we call it a *non-conditional*, or *equational*, QIT, otherwise it is called a *conditional* [12], or *quasi-equational*, QIT. If all of the constant types $A$ in any of the constructors are finite (isomorphic to $\mathsf{Fin}\ n$ for $n : \mathbb{N}$) then it is called a *finitary* QIT [12]. Otherwise, it is called a *generalised* [12], or *infinitary*, QIT. We are not aware of any existing examples in the literature of HITs which allow the point constructors to be conditional (though it is not difficult to imagine), nor any schemes for HITs that allow such definitions. However, we do believe this is worth investigating further.

*Conjecture 1.* Any equational QIT can be encoded as a QW-type.

We believe this can be proved analogously to the approach of Dybjer [11] for inductive types, though the endpoints still need to be considered and we have not yet translated the schema in definition 3 into Agda.

*Remark 3.* Assuming Conjecture 1, Basold, Geuvers, and van der Weide's schema [6], being an equational (non-conditional) instance of Definition 3, can be encoded as a QW-type.

## 4  Construction of QW-types

In Section 2 we defined a QW-type to be initial among algebras over a given (possibly infinitary) signature satisfying a given systems of equations (Definition 2). If one interprets these notions in classical Zermelo-Fraenkel set theory with the axiom of Choice (ZFC), one regains the usual notion from universal algebra of initial algebras for infinitary equational theories. Since in the set-theoretic interpretation there is an upper bound on the cardinality of arities of operators in a given signature $\Sigma$, the ordinal-indexed sequence $\mathsf{S}^\alpha(\varnothing)$ of iterations of the functor in (2) starting from the empty set eventually becomes stationary; and so the sequence has a small colimit, namely the set $\mathsf{W}\{\Sigma\}$ of well-founded trees over $\Sigma$. A system of equations $\varepsilon$ (Definition 1) over $\Sigma$ generates a $\Sigma$-congruence relation $\sim$ on $\mathsf{W}\{\Sigma\}$. The quotient set $\mathsf{W}\{\Sigma\}/\sim$ yields the desired initial algebra for $(\Sigma, \varepsilon)$ provided the $\mathsf{S}$-algebra structure on $\mathsf{W}\{\Sigma\}$ induces one on the quotient set. It does so, because for each operator, using AC one can pick representatives of the (possibly infinitely many) equivalence classes that are the arguments of the operator, apply the interpretation of the operator in $\mathsf{W}\{\Sigma\}$ and then take the equivalence class of that. So the set-theoretic model of type theory in ZFC models QW-types.

Is this use of choice really necessary? Blass [7, Section 9] shows that if one drops AC and just works in ZF, then provided a certain large cardinal axiom is consistent with ZFC, it is consistent with ZF that there is an infinitary equational theory with no initial algebra. He shows this by first exhibiting a countably presented equational theory whose initial algebra has to be an uncountable regular cardinal; and secondly appealing to the construction of Gitik [17] of a model of ZF with no uncountable regular cardinals (assuming a certain large cardinal axiom). Lumsdaine and Shulman [21] turn the infinitary equational theory of Blass into a higher-inductive type that cannot be proved to exist in ZF (and hence cannot be constructed in type theory just using pushouts and the natural numbers). We noted in Example 5 that this higher inductive type can be presented as a QW-type.

So one cannot hope to construct QW-types using a type theory which is interpretable in just ZF. However, the type theory in which we work, with its universes closed under inductive-inductive definitions, already requires going beyond ZF to be able to give it a naive, classical set-theoretic interpretation (by assuming the existence of enough strongly inaccessible cardinals, for example). So the above considerations about initial algebras for infinitary equational theories in classical set theory do not rule out the construction of QW-types in the type theory in which we work. However, something more than just quotienting a W-type is needed in order to prove Theorem 1.

Figure 2 gives a first attempt to do this (which later we will modify using sized types to get around a termination problem). The definition is relative to a given signature $\Sigma : \mathsf{Sig}$ and system of equations $\varepsilon = (E, V, l, r) : \mathsf{Syseq}\, \Sigma$. It makes use of quotient types, which we add to Agda via postulates, as shown in Figure 3.[4] The REWRITE pragma makes $\mathsf{elim}\, R\, B\, f\, e\, (\mathsf{mk}\, R\, x)$ definitionally equal to $f\, x$ and is not merely a computational convenience—this is what allows function extensionality to be proved from these postulated quotient types. The POLARITY pragmas enable the postulated quotients to be used in datatype declarations at positions that Adga deems to be strictly positive; a case in point being the definitions of $\mathsf{Q}_0$ and $\mathsf{Q}_1$ in Figure 2. Agda's test for strict positivity is sound with respect to a set-theoretic semantics of inductively defined datatypes that are built up using strictly positive uses of dependent functions; the semantics of such datatypes uses initial algebras for endofunctors possessing a rank. Here we

---

[4] The actual implementation is polymorphic in universe levels, but for simplicity here we just give the level-zero version.

```
mutual
  data Q₀ : Set where
    sq : T Q → Q₀

  data Q₁ : Q₀ → Q₀ → Set where
    sqeq : (e : E)(ρ : V e → Q) → Q₁ (sq(T'ρ(l e))) (sq(T'ρ(r e)))
    sqη : (x : Q₀) → Q₁ (sq(η(qu x))) x
    sqσ : (s : S(T Q)) → Q₁ (sq(σ s)) (sq(ι(S'(qu ∘ sq) s)))

  Q : Set
  Q = Q₀/Q₁

  qu : Q₀ → Q
  qu = quot.mk Q₁

QW{Σ}{ε} = Q
```

**Figure 2.** First attempt at constructing QW-types

are allowing the inductively defined datatypes to be built up using quotients as well, but this is semantically unproblematic, since quotienting does not increase rank. (Later we need to combine the use of POLARITY with sized types; the semantics of this has been studied for System $F_\omega$ [3], but needs to be explored further for Agda.)

We build up the underlying inductive type $Q_0$ to be quotiented using a constructor sq that takes well-founded trees $T(Q_0/Q_1)$ of whole equivalence classes with respect to a relation $Q_1$ that is mutually inductively defined with $Q_0$—an instance of an inductive-inductive definition [15]. The definition of $Q_1$ makes use of the actions on functions of the signature endofunctor S and its associated free monad T (Section 2); those actions are defined as follows:

$$\begin{aligned}&\mathsf{S'} : \{X\ Y : \mathsf{Set}\} \to (X \to Y) \to \mathsf{S}\,X \to \mathsf{S}\,Y \\ &\mathsf{S'}\,f\,(a,b) = (a, f \circ b)\end{aligned} \quad (24)$$

$$\begin{aligned}&\mathsf{T'} : \{X\ Y : \mathsf{Set}\} \to (X \to Y) \to \mathsf{T}\,X \to \mathsf{T}\,Y \\ &\mathsf{T'}\,f\,t = t \ggg (\eta \circ f)\end{aligned} \quad (25)$$

The definition of $Q_1$ also uses the natural transformation $\iota : \{X : \mathsf{Set}\} \to \mathsf{S}\,X \to \mathsf{T}\,X$ defined by $\iota = \sigma \circ \mathsf{S'}\,\eta$.

Turning to the proof of Theorem 1 using the definitions in Figure 2, the S-algebra structure (9) is easy to define without using any form of choice, because of the type of $Q_0$'s constructor sq. Indeed, we can just take qwintro to be $\mathsf{qu} \circ \mathsf{sq} \circ \iota : \mathsf{S}(\mathsf{QW}) \to \mathsf{QW}$.[5] The first constructor sqeq of the data type $Q_1$ ensures that the quotient $Q_0/Q_1$ satisfies the equations in $\varepsilon$, so that we get qwequ as in (10); and the other two constructors, sqη and sqσ make identifications that

---

[5] The use of the free monad $T\{\Sigma\}$ in the domain of sq, rather than just $S\{\Sigma\}$, seems necessary in order to define $Q_1$ with the properties needed for (10)–(13).

```
module quot where
  postulate
    ty : {A : Set}(R : A → A → Set) → Set
    mk : {A : Set}(R : A → A → Set) → A → ty R
    eq : {A : Set}(R : A → A → Set){x y : A} → R x y → mk R x ≡ mk R y
    elim : {A : Set}(R : A → A → Set)(B : ty R → Set)(f : (x : A) → B(mk R x))
         (e : {x y : A} → R x y → f x ≡≡ f y)(z : ty R) → B z
    comp : {A : Set}(R : A → A → Set)(B : ty R → Set)(f : (x : A) → B(mk R x))
         (e : {x y : A} → R x y → f x ≡≡ f y)(x : A) → elim R B f e (mk R x) ≡ f x
{-# REWRITE comp -#}
{-# POLARITY ty ++ ++ -#}
{-# POLARITY mk _ _ * -#}

_/_ : (A : Set)(R : A → A → Set) → Set
A/R = quot.ty R
```

**Figure 3.** Quotient types

enable the construction of functions qwrec, qwrechom and qwuniq as in (11)–(13). However, there is a problem. Given $X$ : Set, $s$ : S $X \to X$ and $e$ : Sat $X$, for qwrec $X\,s\,e$ we have to construct a function r : Q → $X$. Since Q = $Q_0/Q_1$ is a quotient, we will have to use the eliminator quot.elim from Figure 3 to define r. The following is an obvious candidate definition

$$\begin{aligned}
&\text{mutual} \\
&\quad \text{r} : \text{Q} \to X \\
&\quad \text{r} = \text{quot.elim}\,Q_1\,(\lambda\_\to X)\,r_0\,r_1 \\
&\quad r_0 : Q_0 \to X \\
&\quad r_0(\text{sq}\,t) = t \ggg= \text{r} \\
&\quad r_1 : \{x\,y : Q_0\} \to Q_1\,x\,y \to r_0\,x \equiv r_0\,y \\
&\quad r_1 = \cdots
\end{aligned} \tag{26}$$

(where we have elided the details of the invariance proof $r_1$). The problem with this mutually recursive definition is that it is not clear to us (and certainly not to Agda) whether it gives totally defined functions: although the value of $r_0$ at a typical element sq $t$ is explained in terms of the structurally smaller element $t$, the explanation involves r, whose definition uses the whole function $r_0$ rather than some application of it at a structurally smaller argument. Agda's termination checker rejects the definition.

We get around this problem by using a type-based termination method, namely Agda's implementation of sized types [2]. Intuitively, this provides a type Size of "sizes" which give a constructive abstraction of features of ordinals in ZF when they are used to index sequences of sets that eventually become stationary, such as in various transfinite constructions of free algebras [20, 14]. In Agda, the type Size comes equipped with various relations and functions: given sizes

```
mutual
  data Q₀(i : Size) : Set where
    sq : {j : Size< i} → T(Q j) → Q₀ i

  data Q₁(i : Size) : Q₀ i → Q₀ i → Set where
    sqeq : {j : Size< i}(e : E)(ρ : V e → Q j) → Q₁ i (sq(T'ρ (l e))) (sq(T'ρ (r e)))
    sqη : {j : Size< i}(x : Q₀ j) → Q₁ i (sq(η(qu j x))) (φ₀ i x)
    sqσ : {j : Size< i}{k : Size< j}(s : S(T(Q k))) →
               Q₁ i (sq(σ s)) (sq(ι(S'(qu j ∘ sq) s)))

  Q : Size → Set
  Q i = (Q₀ i)/Q₁ i

  qu : (i : Size) → Q₀ i → Q i
  qu i = quot.mk (Q₁ i)

  φ₀ : (i : Size){j : Size< i} → Q₀ j → Q₀ i
  φ₀ i (sq z) = sq z

QW{Σ}{ε} = Q ∞
```

**Figure 4.** Construction of QW-types using sized types

$i, j$ : Size, there is a relation $i$ : Size$< j$ to indicate strictly increasing size (so the type Size$< j$ is treated as a subtype of Size); there is a successor operation ↑ : Size → Size (and also a join operation $\_\sqcup^s\_$ : Size → Size → Size, but we do not need it here); and a size $\infty$ : Size to indicate where a sequence becomes stationary. Thus we construct the QW-type QW$\{\Sigma\}\{\varepsilon\}$ as Q $\infty$ for a suitable size-indexed sequence of types Q : Size → Set, shown in Figure 4.

For each size $i$ : Size, the type Q $i$ is a quotient Q$_0$ $i$/Q$_1$ $i$, where the constructors of the data types Q$_0$ $i$ and Q$_1$ $i$ take arguments of smaller sizes $j$ : Size$< i$. Consequently in the following sized version of (26)

```
mutual                                                                    (27)
  r : {i : Size} → Q i → X
  r{i} = quot.elim (Q₁ i) (λ_ → X) (r₀ {i}) (r₁ {i})

  r₀ : {i : Size} → Q₀ i → X
  r₀{i}(sq {j} t) = t ⋙= r {j}

  r₁ : {i : Size}{x y : Q₀ i} → Q₁ i x y → r₀ x ≡ r₀ y
  r₁ = ⋯
```

the definition of r$_0\{i\}$ involves a recursive call via r to the whole function r$_0$, but at a size $j$ which is smaller than $i$. So now Agda accepts that the definition of qwrec $X s e$ as r $\infty$, with r as in (27), is terminating.

Thus we get a function qwrec for (11). We still have (9), but now with qwintro = qu $\infty$ ∘ sq $\{\infty\}$ ∘ ι; and as before, the constructor sqeq of Q$_1$ in Figure 4 ensures that QW = (Q$_0$ $\infty$)/Q$_1$ $\infty$ satisfies the equations $\varepsilon$. With these definitions it turns out that each qwrec $X s e$ is an S-algebra morphism up to definitional equality, so that the function qwrechom needed for (12) is straightforward to define. Finally, the function qwuniq needed for (13) can be constructed via a sequence of lemmas making use of the other two constructors of the data type $Q_1$, namely sq$\eta$, which makes use of an auxiliary function for coercing between different size instances of $Q_0$, and sq$\sigma$. We refer the reader to the accompanying Agda code (DOI: 10.17863/CAM.48187) for the details of the construction of qwuniq. Altogether, the sized definitions in Figure 4 allow us to complete a proof of Theorem 1.

## 5   Conclusion

QW-types are a general form of QIT that capture many examples, including simple 1-cell complexes and non-recursive QITs [6], non-structural QITs [26], W-types with reductions [28], and also infinitary QITs (e.g. unordered infinitely branching trees [5], and ordinals [21]). They also capture the notion of initial (and free) algebras for strictly positive equational systems [14], analogously to how W-types capture the notion of initial (and free) algebras for strictly positive endofunctors (see Remark 2). Using Agda to formalise our results, we have shown that it is possible to construct any QW-type, even infinitary ones, in intensional type theory satisfying UIP, using inductive-inductive definitions permitting strictly positive occurrences of quotients and sized types (see Theorem 1 and Section 4). We conclude by mentioning related work and some possible directions for future work.

*Quotients of monads.* In view of Remark 2, Section 4 gives a construction of initial algebras for equational systems [14] on the *free* monad $\mathsf{T}\{\Sigma\}$ generated by a signature $\Sigma$. By a suitable change of signature (see Remark 1) this extends to a construction of free algebras, rather than just initial ones. We can show that the construction works for an arbitrary strictly positive monad and not just for free ones. Given such a construction one gets a quotient monad morphism from the base monad to the quotient monad. This contravariantly induces a forgetful functor from the algebras of the latter to that of the former. Using the adjoint triangle theorem, one should be able to construct a left adjoint. This would then cover examples such as the free group over a monoid, free ring over a group, etc.

*Quotient inductive-inductive types.* The notion of QW-type generalises to *indexed* QW-types, analogously to the generalisation of W-types to Petersson-Synek trees for inductively defined indexed families of types [24, Chapter 16], and we will consider it in subsequent work. More generally, we wonder whether our analysis of QITs using quotients, inductive-inductive and sized types can be extended to cover the notion of *quotient inductive-inductive* type (QIIT) [4, 19]. Dijkstra [10] studies such types in depth and in Chapter 6 of his thesis gives a construction for finitary ones in terms of countable colimits, and hence in terms of countable coproducts and quotients. One could hope to pass to the infinitary case by using sized types as we have done, provided an analogue for QIITs can be found of the monadic construction in Section 4 for our class of QITs, the QW-types. Kaposi, Kovács, and Altenkirch [19] give a specification of finitary QIITs using a domain-specific type theory called the *theory of signatures* and prove existence of QIITs matching this specification. It might be possible to encode their theory of signatures using QW-types (it can already be encoded as a QIIT), or to extend QW-types making this possible. This would allow infinitary QIITs.

*Schemas for QITs.* We have shown by example that QW-types can encode a wide range of QITs. However, we have yet to extend this to a proof of Conjecture 1 that every instance of the schema for QITs considered in Section 3 can be so encoded.

*Conditional path equations.* In Section 3 we mentioned the fact that Dybjer and Moeneclaey [12] give a model for finitary 1-HITs and 2-HITs in which constructors are allowed to take arguments involving the identity type of the datatype being declared. On the face of it, QW-types are not able to encode such *conditional* QITs. We plan to consider whether it is possible to extend the notion of QW-type to allow encoding of infinitary QITs with such conditional equations.

*Homotopy Type Theory (HoTT).* Our development makes use of UIP (and heterogeneous equality), which is well-known to be incompatible with the Univalence Axiom [29, Example 3.1.9]. Given the interest in HoTT, it is certainly worth investigating whether a result like Theorem 1 holds in univalent foundations for a suitably coherent version of QW-types. We are currently investigating this using set-truncation.

*Pattern matching for QITs and HITs.* Our reduction of QITs to induction-induction, strictly positive quotients and sized types is of theoretical interest, but in practice one could wish for more direct support in systems like Agda, Lean and Coq for the very useful notion of quotient inductive types (or more generally, for higher inductive types). Even having better support for the special case of quotient types would be welcome. It is not hard to envisage the addition of a general schema for declaring QITs; but when it comes to defining functions on them, having to do that with eliminator forms rapidly becomes cumbersome (for example, for functions of several QIT arguments). Some extension of dependently typed pattern matching to cover equality constructors as well as element constructors is needed and the third author has begun work on that based on the approach of Cockx and Abel [9].[6]

---

[6] In this context it is worth mentioning that the `cubical` features of recent versions of Agda give access to cubical type theory [30]. This allows for easy declaration of HITs and hence in particular QITs (and quotients avoiding the need for POLARITY pragmas) and a certain amount of pattern matching when it comes to defining functions on them: the value of a function on a path constructor can be specified by using generic elements of the interval type in point-level patterns; but currently the user is given little mechanised assistance to solve the definitional equality constraints on end-points of paths that are generated by this method.